\documentclass[
reprint,           
superscriptaddress,
amsmath,           
amssymb,           
aps,               
prd,               
notitlepage,       
longbibliography,  
floatfix,          
nofootinbib,
]{revtex4-1}
\usepackage[
colorlinks=true,        
citecolor=blue,         
linkcolor=blue,         
urlcolor=blue           
]{hyperref}       
\usepackage{amsmath}
\usepackage{graphicx}   

\newcommand{\nc}{\newcommand*} 

\nc{\al}{\alpha}
\nc{\s}{\sigma}
\nc{\dt}{\delta}
\nc{\Dt}{\Delta}
\nc{\Ld}{\Lambda}
\nc{\p}{\partial}
\nc{\om}{\omega}
\nc{\Om}{\Omega}
\nc{\rd}{\mathrm{d}}
\nc{\Od}[1]{\mathcal{O}(#1)} 
\nc{\kp}{\kappa}

\def\({\left(}
\def\){\right)}
\def\[{\left[}
\def\]{\right]}
\def\e{\begin{equation}}
\def\q{\end{equation}}
\def\m{\begin{eqnarray}}
\def\n{\end{eqnarray}}

\nc{\Eq}[1]{Eq.~\eqref{#1}}     
\nc{\Fig}[1]{Fig.~\ref{#1}}     
\nc{\Table}[1]{Table~\ref{#1}}  
\nc{\Sec}[1]{Sec.~\ref{#1}}     

\nc{\Msun}{M_\odot}             
\nc{\fpbh}{f_{\mathrm{pbh}}}    
\nc{\fpbhn}{f_{\mathrm{pbh0}}}    
\nc{\mR}{\mathcal{R}} 
\nc{\seq}{\sigma_{\mathrm{eq}}}
\nc{\ogw}{\Omega_{\mathrm{GW}}}
\nc{\gpcyr}{\mathrm{Gpc}^{-3}\,\mathrm{yr}^{-1}}
\nc{\lvc}{LIGO/Virgo} 
\nc{\SNR}{\mathrm{SNR}} 
\nc{\mmin}{{m_{\mathrm{min}}}}
\nc{\mmax}{{m_{\mathrm{max}}}}
\nc{\Mmin}{{M_{\mathrm{min}}}}
\nc{\fmin}{{f_{\mathrm{min}}}}
\nc{\VT}{\mathrm{VT}}
\nc{\rhoGW}{\rho_{\mathrm{GW}}}
\nc{\vth}{\vec{\theta}}
\nc{\vd}{\vec{d}}
\nc{\vla}{\vec{\lambda}}
\nc{\Nobs}{N_{\mathrm{obs}}}
\nc{\av}[1]{\langle #1 \rangle} 
\nc{\km}{\mathrm{km}}
\nc{\Mpc}{\mathrm{Mpc}}
\nc{\Tobs}{T_{\mathrm{obs}}}
\nc{\Ntemp}{N_{\mathrm{temp}}}

\nc{\addref}{[\textcolor{red}{add ref}] } 
\nc{\eg}{\textit{e.g.~}}
\nc{\app}{\approx}
\nc{\hf}{\frac{1}{2}}
\nc{\discuss}{\textcolor{red}{Add discussion here!}}

\nc{\mpbh}{m_{\rm{pbh}}}
\nc{\cR}{\mathcal{R}}
\nc{\mU}{{\mathcal{U}}}
\nc{\Mc}{{M_\mathrm{c}}}
\nc{\Mf}{{M_\mathrm{f}}}
\nc{\red}[1]{\textcolor{red}{#1}}
\nc{\yellow}[1]{\textcolor{yellow}{#1}}
\nc{\green}[1]{\textcolor{green}{#1}}
\nc{\blue}[1]{\textcolor{blue}{#1}}

\begin{document}
	
\title{Constraining the Merger History of Primordial-Black-Hole Binaries from GWTC-3}
	
\author{Lang~Liu}
\email{liulang@bnu.edu.cn}
\affiliation{Department of Astronomy, Beijing Normal University, Beijing 100875, China}
\affiliation{Advanced Institute of Natural Sciences, Beijing Normal University, Zhuhai 519087, China}

\author{Zhi-Qiang~You}
\email{Corresponding author: zhiqiang.you@bnu.edu.cn}
\affiliation{Department of Astronomy, Beijing Normal University, Beijing 100875, China}
\affiliation{Advanced Institute of Natural Sciences, Beijing Normal University, Zhuhai 519087, China}

\author{You~Wu}
\email{youwuphy@gmail.com}
\affiliation{College of Mathematics and Physics, Hunan University of Arts and Science, Changde, 415000, China}

\author{Zu-Cheng~Chen}
\email{Corresponding author: zucheng.chen@bnu.edu.cn}
\affiliation{Department of Astronomy, Beijing Normal University, Beijing 100875, China}
\affiliation{Advanced Institute of Natural Sciences, Beijing Normal University, Zhuhai 519087, China}

\begin{abstract}
Primordial black holes (PBHs) can be not only cold dark matter candidates but also progenitors of binary black holes observed by LIGO-Virgo-KAGRA (LVK) Collaboration. The PBH mass can be shifted to the heavy distribution if multi-merger processes occur. In this work, we constrain the merger history of PBH binaries using the gravitational wave events from the third Gravitational-Wave Transient Catalog (GWTC-3). Considering four commonly used PBH mass functions, namely the log-normal, power-law, broken power-law, and critical collapse forms, we find that the multi-merger processes make a subdominant contribution to the total merger rate. Therefore, the effect of merger history can be safely ignored when estimating the merger rate of PBH binaries. We also find that GWTC-3 is best fitted by the log-normal form among the four PBH mass functions and confirm that the stellar-mass PBHs cannot dominate cold dark matter.
\end{abstract}
	
\maketitle

\section{Introduction}

The successful detection of gravitational waves (GWs) from compact binary coalescences \cite{LIGOScientific:2018mvr,LIGOScientific:2020ibl,LIGOScientific:2021djp} has led us into a new era of GW astronomy.
According to the recently released third GW Transient Catalog (GWTC-3) \cite{LIGOScientific:2021djp} by LIGO-Virgo-KAGRA (LVK) Collaboration, there are $90$ GW events detected during the first three observing runs. Most of these events are categorized as binary black hole (BBH) mergers, and the BBHs detected by LVK have a broad mass distribution. The heaviest event, GW190521 \cite{LIGOScientific:2020iuh}, has component masses $m_1 = 85^{+21}_{-14}\Msun$ and $m_2 = 66^{+17}_{-18}\Msun$. Both masses lie within upper black hole mass gap originated from pulsation pair-instability supernovae \cite{Marchant_2019}, and current modelling places the lower cutoff of the mass gap at $\sim 50\pm 4\Msun$ \cite{Belczynski:2016jno,Marchant_2019,Farmer_2019,Farmer:2020xne,Marchant:2020haw}.
Even accounting for the statistical uncertainties, it still implies at least $m_1$ is well within the mass gap and cannot originate directly from a stellar progenitor \cite{Anagnostou:2020tta}.
Therefore, the heavy event GW190521 greatly challenges the stellar evolution scenario of astrophysical black holes.

Besides the astrophysical black holes, another possible explanation for the LVK BBHs is the primordial black holes (PBHs) \cite{Bird:2016dcv,Sasaki:2016jop,Chen:2018czv,Liu:2018ess,Chen:2021nxo}. 
PBHs are black holes formed in the very early Universe through the gravitational collapse of the primordial density fluctuations \cite{Hawking:1971ei,Carr:1974nx}. Recently, PBHs have attracted considerable attention \cite{Garcia-Bellido:2017mdw,Carr:2017jsz,Germani:2017bcs,Liu:2019rnx,Cai:2019elf,Cai:2019bmk,DeLuca:2020sae,Vaskonen:2020lbd,DeLuca:2020agl,Hutsi:2020sol,Sasaki:2018dmp,Carr:2020gox,Carr:2020xqk,Liu:2021jnw,Wang:2022nml,Franciolini:2022tfm} because they can be not only the sources of LVK detections~\cite{Bird:2016dcv,Sasaki:2016jop}, but also candidates of cold dark matter (CDM) \cite{Carr:2016drx} and the seeds for galaxy formation~\cite{Bean:2002kx,Kawasaki:2012kn}. The formation of PBHs would inevitably accompany the production of scalar-induced GWs \cite{Saito:2008jc,Cai:2018dig,Yuan:2019udt,Yuan:2019wwo,Yuan:2019fwv,Chen:2019xse,DeLuca:2019ufz,Bartolo:2018rku,Bartolo:2018evs}.
Recent studies \cite{Chen:2021nxo,Chen:2022fda} show that the BBHs from GWTC-3 are consistent with the PBH scenario, and the abundance of PBH in CDM, $\fpbh$, should be in the order of $\mathcal{O}(10^{-3})$ to explain LVK BBHs. In particular, the merger rate for GW190521 derived from the PBH model is consistent with that inferred by LVK, indicating that GW190521 can be a PBH binary \cite{DeLuca:2020sae,Chen:2021nxo}.

Accurately estimating the merger rate distribution of PBH binaries can be crucial to extract the PBH population parameters from GW data. Ref.~\cite{Liu:2019rnx} studies the multi-merger processes of PBH binaries and show that the merger history of PBH binaries may shift the mass distribution from light mass to heavy mass depending on the values of population parameters. Ref.~\cite{Wu:2020drm} then infers the population parameters of PBH binaries by accounting for the merger history effect using $10$ BBHs from GWTC-1, finding that the effect of merger history can be safely ignored when estimating the merger rate of PBH binaries. 
In this work, we use the LVK recent released GWTC-3 data to constrain the effect of merger history on the merger rate of PBH binaries assuming all LVK BBHs are of primordial origin.
We extend the analyses of Ref.~\cite{Wu:2020drm} in several aspects. Firstly, we use a purified subset of GWTC-3, which expands GWTC-1 with almost six times more BBH events. The GWTC-3 events expand the mass and redshift coverage and can alleviate the statistical bias by including significantly more BBHs. Secondly, Ref.~\cite{Wu:2020drm} only considers the PBH mass functions with the log-normal and power-law forms. We do more comprehensive analyses by including the broken power-law and critical collapse PBH mass functions that were not considered in Ref.~\cite{Wu:2020drm}. It is claimed by Ref.~\cite{Deng:2021ezy} that a broken power-law can fit the GW data better than the log-normal form. Lastly, we consider the redshift distribution of the merger rate that is ignored in Ref.~\cite{Wu:2020drm}.
The aforementioned reasons have inspired us to explore the possibility that the heavy black holes detected by LVK have been formed, at least in part, through second-generation mergers. This is because the second-merger process has the potential to increase the mass distribution to a higher value. A precise assessment of the influence of second-generation mergers on mass distribution demands a meticulous analysis of the data, as has been conducted in this study.

We organize the rest paper as follows.
In \Sec{merger}, we briefly review the calculation of the merger rate of PBH binaries by accounting for the merger history effect.
In \Sec{method}, we describe the hierarchical Bayesian framework used to infer the PBH population parameters from GW data.
In \Sec{result}, we consider four commonly used PBH mass functions and present the results.
Finally, we give conclusions in \Sec{conclusion}.

\section{\label{merger}Merger rate density distribution of PBH binaries}

In this section, we will outline the calculation of merger rate density when considering the PBH merger history effect. We refer to Ref.~\cite{Liu:2019rnx} for more details.

The BBHs observed by LVK suggest that BHs should have a broad mass distribution, so we consider an extended mass function for PBHs. Here, we demand the probability distribution function of PBH mass, $P(m)$, be normalized such that
\e\label{norm}
\int_{0}^{\infty} P(m)\, \rd m = 1.
\q 
Assuming the fraction of PBHs in CDM is $\fpbh$, we can estimate the abundance of PBHs in the mass interval $(m, m+\rd m)$ as \cite{Chen:2018rzo}
\e 
0.85 \fpbh\, P(m)\, \rd m.
\q 
The coefficient $0.85$ is roughly the fraction of CDM in the non-relativistic matter, including both CDM and baryons. Following Ref.~\cite{Liu:2019rnx}, we may define an average PBH mass, $\mpbh$, as
\e\label{mpbh}
\frac{1}{\mpbh} = \int \frac{P(m)}{m} \rd m.
\q 
Then, we can obtain the average number density of PBHs with mass $m$ in the 
total number density of PBHs, $F(m)$, by \cite{Liu:2019rnx}
\e\label{Fm} 
F(m) = P(m) \frac{\mpbh}{m}.
\q 

We can now estimate the merger rate densities of PBH binaries by considering the merger history effect.
We assume that PBHs are randomly distributed following a spatial Poisson distribution in the early Universe when they decouple from the cosmic background evolution \cite{Nakamura:1997sm,Sasaki:2016jop,Ali-Haimoud:2017rtz}. The two nearest PBHs would attract each other because of the gravitational interactions. These two PBHs would obtain the angular momentum from the torque of other PBHs and form a PBH binary after decoupling from the cosmic expansion. The binary would emit gravitational radiations and eventually merge. 

We do not intend to give a detailed derivation but quote the results from Ref.~\cite{Liu:2019rnx} here.
The merger rate density from first-merger process, $\cR_1(t, m_i, m_j)$, is given by \cite{Liu:2019rnx}
\e 
\cR_1(t, m_i, m_j) = \int \hat{\cR}_1 \rd m_l,
\q 
where $m_i$ and $m_j$ are the masses of the merging binary, $m_l$ is the mass of the third black hole that is closest to the merging binary, and
\e 
\begin{split}
	&\hat{\cR}_1(t, m_i, m_j, m_l)
	\equiv 1.32 \times 10^6 \times \(\frac{t}{t_0}\)^{-\frac{34}{37}}\(\frac{\fpbh}{\mpbh}\)^\frac{53}{37} \\ 
	&\times m_l^{-\frac{21}{37}} (m_i m_j)^\frac{3}{37} (m_i + m_j)^\frac{36}{37} F(m_i) F(m_j) F(m_l).     
\end{split}
\q
Here, $t$ is the cosmic time, and $t_0$ is the present cosmic time.
Similarly, the merger rate density from second-merger process, $\cR_2(t, m_i, m_j)$, is given by \cite{Liu:2019rnx}
\e 
\begin{split}
	\cR_2(t, m_i, m_j) &= \hf \int \hat{\cR}_2(t, m_i-m_e, m_e, m_j, m_l)\ \rd m_l \rd m_e\\
	&+ \hf \int \hat{\cR}_2(t, m_j-m_e, m_e, m_i, m_l)\ \rd m_l \rd m_e,
\end{split}
\q 
where $m_e$ is the mass of the fourth black hole that is closest to the merging binary, and
\e 
\begin{split}
	&\hat{\cR}_2(t, m_i, m_j,m_k, m_l)
	=1.59 \times 10^4 \times \(\frac{t}{t_0}\)^{-\frac{31}{37}} \(\frac{\fpbh}{\mpbh}\)^\frac{69}{37}\\ 
	&\qquad\times m_k^{\frac{6}{37}} m_l^{-\frac{42}{37}} (m_i+ m_j)^\frac{6}{37} (m_i + m_j + m_k)^\frac{72}{37}\\
	&\qquad\times F(m_i) F(m_j) F(m_k) F(m_l).     
\end{split}
\q
We only consider the effect of merger history up to the second-merger process. We have verified that the fraction of the third merger rate over the second merger rate is less than $0.005$.
Therefore the total merger rate density, $\cR(t, m_i, m_j)$, of PBH binaries at cosmic time $t$ with masses $m_i$ and $m_j$ is
\e\label{cR}
\cR(t, m_i, m_j) =  \sum_{n=1,2} \cR_n(t, m_i, m_j),
\q 
and the total merger rate is
\e\label{Rt}
R(t) = \int \cR(t, m_i, m_j) \rd m_i \rd m_j = \sum_{n=1,2} R_n(t),
\q 
where
\e 
R_n(t) = \int \cR_n(t, m_i, m_j) \rd m_i \rd m_j.
\q 
All the above-mentioned merger rate (density) is measured at the source frame.
We should emphasize that although $R_2(t)$ should be smaller than $R_1(t)$ as expected, $\cR_2(t, m_i, m_j)$ is not necessarily be smaller than $\cR_1(t, m_i, m_j)$ \cite{Liu:2019rnx}. 

\section{\label{method}Hierarchical Bayesian Inference}
We adopt a hierarchical Bayesian approach to infer the population parameters by marginalizing the uncertainty in estimating individual event parameters.
This section describes the hierarchical Bayesian inference used in the parameter estimations.
The merger rate density \eqref{cR} is measured in the source frame, and we need to convert it into the detector frame as
\e\label{Rpop}
\mR_{\mathrm{pop}}(\theta|\Ld) = \frac{1}{1+z} \frac{dV_\mathrm{c}}{dz} \mR(\theta|\Ld),
\q
where $z$ is the cosmological redshift, $\theta\equiv\{z, m_1, m_2\}$, $\Ld$ is a collection of $\fpbh$ and the parameters from mass function $P(m)$, and $dV_\mathrm{c}/dz$ is the differential comoving volume. The factor $1/(1 + z)$ converts time increments from the source to the detector frame. We take the cosmological parameters from Planck 2018 \cite{Planck:2018vyg}.

\begin{table}[tbp!]
	\centering
	\begin{tabular}{lll}
		\hline\hline
		\textbf{Parameter\quad} & \textbf{Description} & \textbf{Prior} \\
		\hline
		$\fpbh$ & Abundance of PBH in CDM & log-$\mU(-4, 0)$\\
		\hline
		\multicolumn{3}{c}{Lognormal PBH mass function} \\[1pt]
		$\Mc$ & Central mass in $\Msun$. & $\mU(5, 50)$\\
		$\sigma$ & Mass width. & $\mU(0.1, 2)$\\
		\hline
		\multicolumn{3}{c}{Power-law PBH mass function} \\[1pt]
		$\Mmin$ & Lower mass cut-off in $\Msun$. & $\mU(3, 10)$\\
		$\al$ & Power-law index. & $\mU(1.05, 4)$\\
		\hline
		\multicolumn{3}{c}{Broken Power-law PBH mass function} \\[1pt]
		$m_*$ & Peak mass in $\Msun$. & $\mU(5, 50)$\\
		$\al_1$ & First power-law index. & $\mU(0, 3)$\\
		$\al_2$ & Second power-law index. & $\mU(1, 10)$\\
		\hline
		\multicolumn{3}{c}{Critical collapse (CC) PBH mass function} \\[1pt]
		$\Mf$ & Horizon mass scale in $\Msun$. & $\mU(1, 50)$\\
		$\al$ & Universal exponent. & $\mU(0, 5)$\\
		\hline
	\end{tabular}	
	\caption{\label{table:priors}Parameters and their prior distributions used in the Bayesian parameter estimations. Here, $\mU$ and log-$\mU$ denote uniform and log-uniform distributions, respectively.}
\end{table}

Given the data, $\textbf{d} = \{d_1, d_2, \cdots, d_{N_{\mathrm{obs}}}\}$, of $N_{\mathrm{obs}}$ BBH merger events, we model the total number of events as an inhomogeneous Poisson process, yielding the likelihood \cite{Loredo:2004nn,Thrane:2018qnx,Mandel:2018mve}
\begin{equation}\label{L1}
	\mathcal{L}(\textbf{d}|\Ld) \propto N_{\mathrm{exp}}^{N_{\mathrm{obs}}} e^{-N_{\mathrm{exp}}} \prod_{i=1}^{N_{\mathrm{obs}}} \frac{\int \mathcal{L} (d_{i}| \theta)\, \cR_{\mathrm{pop}}(\theta|\Ld) d \theta}{\xi(\Ld)},
\end{equation}
where $N_{\exp}\equiv N_{\exp}(\Ld)$ is the expected number of detections over the timespan of observation. Here $\mathcal{L} (d_{i}|\theta)$ is the individual event likelihood for the $i$th GW event that can be derived from the individual event's posterior by reweighing with the prior on $\theta$. Here, $\xi(\Ld)$ quantifies selection biases for a population with parameters $\Ld$ and is defined by
\e\label{xi}
\xi(\Ld) = \int P_{\mathrm{det}}(\theta)\, \cR_{\mathrm{pop}}(\theta|\Ld)\, \mathrm{d} \theta,
\q
where $P_{\text{det}}(\theta)$ is the detection probability that depends on the source parameters $\theta$.
In practice, we use the simulated injections \cite{ligo_scientific_collaboration_and_virgo_2021_5546676} to estimate $\xi(\Ld)$, and \Eq{xi} can be approximated by a Monte Carlo integral over found injections \cite{KAGRA:2021duu}
\begin{equation}
	\xi(\Ld) \approx \frac{1}{N_{\mathrm{inj}}} \sum_{j=1}^{N_{\text {found }}} \frac{\cR_{\mathrm{pop}}(\theta_{j} | \Ld)}{p_{\mathrm{draw}}(\theta_j)},
\end{equation}
where $N_{\text{inj}}$ is the total number of injections, $N_{\text{found}}$ is the number of successfully detected injections, and $p_{\mathrm{draw}}$ is the probability density function from which the injections are drawn.
Using the posterior samples from each event, we estimate the hyper-likelihood \eqref{L1} as
\begin{equation}\label{L2}
	\mathcal{L}(\textbf{d}|\Ld) \propto N_{\mathrm{exp}}^{N_{\mathrm{obs}}} e^{-N_{\mathrm{exp}}} \prod_{i=1}^{N_{\mathrm{obs}}} \frac{1}{\xi(\Ld)} \left\langle \frac{\cR_{\mathrm{pop}}(\theta|\Ld)}{d_L^2(z)} \right\rangle,
\end{equation}
where $\langle\cdots\rangle$ denotes the weighted average over posterior samples of $\theta$. The denominator $d_L^2(z)$ is the standard priors used in the LVK analysis of individual events where $d_L$ is the luminosity distance.

In this work, we incorporate the PBH population distribution \eqref{cR} into the \texttt{ICAROGW} \cite{Mastrogiovanni:2021wsd} package to estimate the likelihood function \eqref{L2}, and use \texttt{dynesty} \cite{Speagle:2019ivv} sampler called from \texttt{Bilby} \cite{Ashton:2018jfp,Romero-Shaw:2020owr} to sample over the parameter space. We use the GW events from GWTC-3 by discarding events with false alarm rate larger than 1 yr$^{-1}$ and events with the secondary component mass smaller than $3\Msun$ to avoid contamination from putative events involving neutron stars following Ref.~\cite{DeLuca:2021wjr}. A total of $69$ GW events from GWTC-3 meet these criteria and the posterior samples of these BBHs are publicly available from Ref.~\cite{ligo_scientific_collaboration_and_virgo_2021_5655785}.

\section{\label{result}results}

Based on the hierarchical statistical framework, we do the parameter estimations for four different PBH mass functions commonly used in the literature. These mass functions are the log-normal, power-law, broken power-law, and critical collapse (CC) distributions, respectively. We summarize the parameters and their prior distributions in \Table{table:priors}. Below we show the results for each of the PBH mass functions.

\subsection{Log-normal mass function}

\begin{figure}[tbp!]
	\centering
	\includegraphics[width=\linewidth]{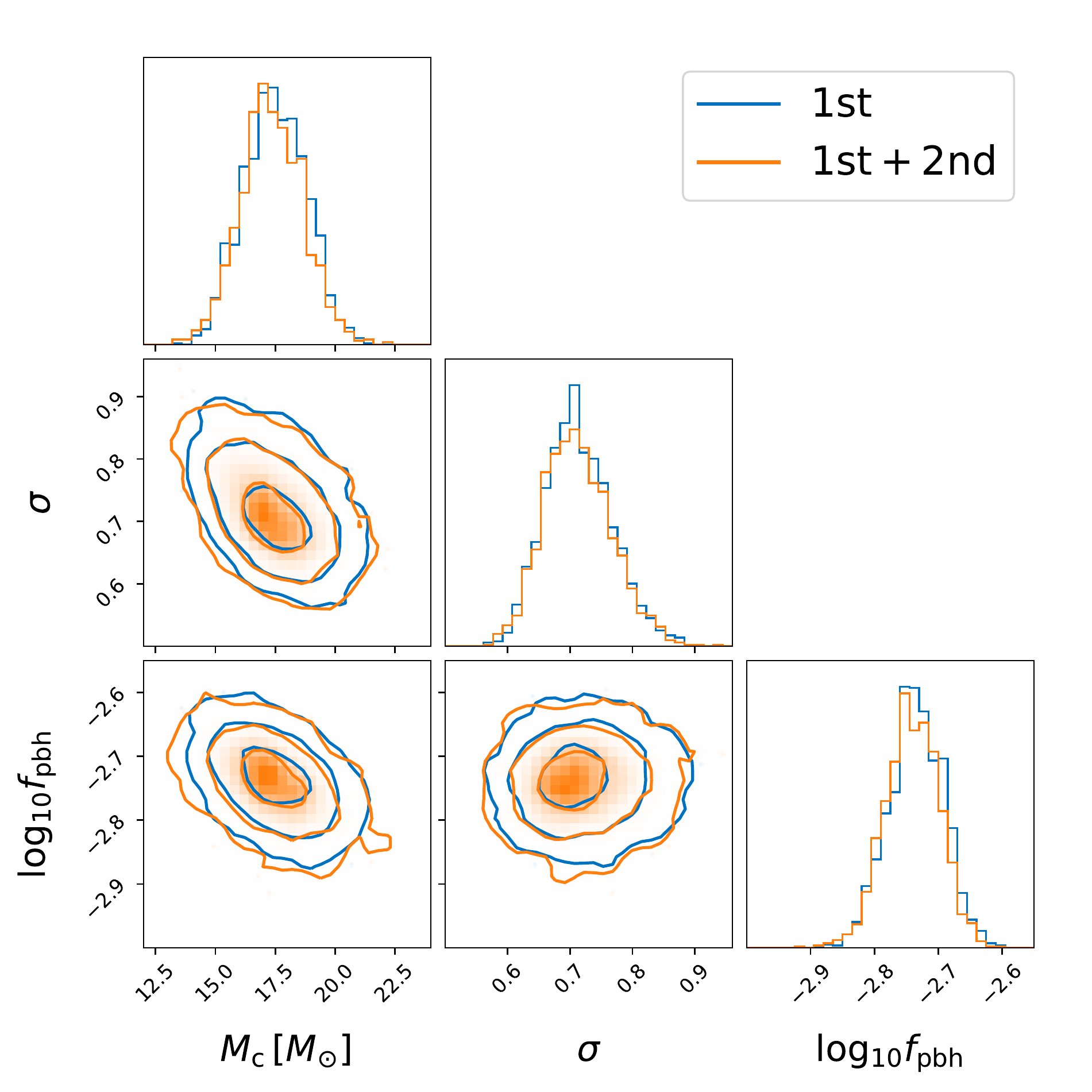}
	\caption{\label{posterior-log}The marginalized one- and two-dimensional posterior distributions for hyper-parameters $\{M_c, \s, \fpbh\}$ in the log-normal mass function inferred from GWTC-3. The blue color denotes the results from the first merger only, while the orange denotes the results from both the first and second mergers. The contours represent the $1\sigma$, $2\sigma$, and $3\sigma$ credible regions, respectively.}
\end{figure}

We first consider a PBH mass function with the log-normal form of \cite{Dolgov:1992pu}
\e\label{log}
P(m) = \frac{1}{\sqrt{2\pi} \s m} \exp \(-\frac{\ln^2\(m/M_c\)}{2\s^2}\),
\q
where $M_c$ represents the central mass of $m P(m)$, and $\s$ characterizes the width of the mass spectrum.
The log-normal mass function can approximate a huge class of extended mass distributions if PBHs are formed from a smooth, symmetric peak in the inflationary power spectrum when the slow-roll approximation holds \cite{Green:2016xgy,Carr:2017jsz,Kannike:2017bxn}.
The hyper-parameters are $\Ld = \{M_c, \s, \fpbh\}$ in this case. 
We can then derive the averaged PBH mass and averaged number density from \Eq{mpbh} and \Eq{Fm} as
\e
\mpbh = M_c \exp \(-\frac{\s^2}{2}\),
\q
\e 
F(m) = \frac{M_c}{\sqrt{2\pi} \s m^2} \exp \(-\frac{\s^2}{2}-\frac{\ln^2\(m/M_c\)}{2\s^2}\).
\q

Using $69$ BBHs from GWTC-3 and performing the hierarchical Bayesian inference, we obtain $M_c = 17.3^{+2.2}_{-2.0} \Msun$, $\s = 0.71^{+0.10}_{-0.08}$, and $\fpbh = 1.8^{+0.3}_{-0.3} \times 10^{-3}$. In this work, we present results with median value and 90\% equal-tailed credible intervals. The posteriors for the hyper-parameters $\Ld = \{M_c, \s, \fpbh\}$ are shown in \Fig{posterior-log}. Note that we get a larger value of $M_c$ than that inferred from GWTC-1 in Ref.~\cite{Wu:2020drm} because GWTC-3 contains heavier BHs than those from GWTC-1. From \Eq{Rt}, we also infer the local merger rate as $R(t_0) = 41^{+16}_{-12} \gpcyr$. The results of local merger rate and abundance of PBHs are consistent with the previous estimations \cite{Sasaki:2016jop,Ali-Haimoud:2017rtz,Chen:2018czv,Chen:2018rzo,Chen:2019irf,Wu:2020drm,Chen:2021nxo,Chen:2022fda,Zheng:2022wqo}, confirming that CDM cannot be dominated by the stellar-mass PBHs.

\begin{figure}[tbp!]
	\centering
	\includegraphics[width=\linewidth]{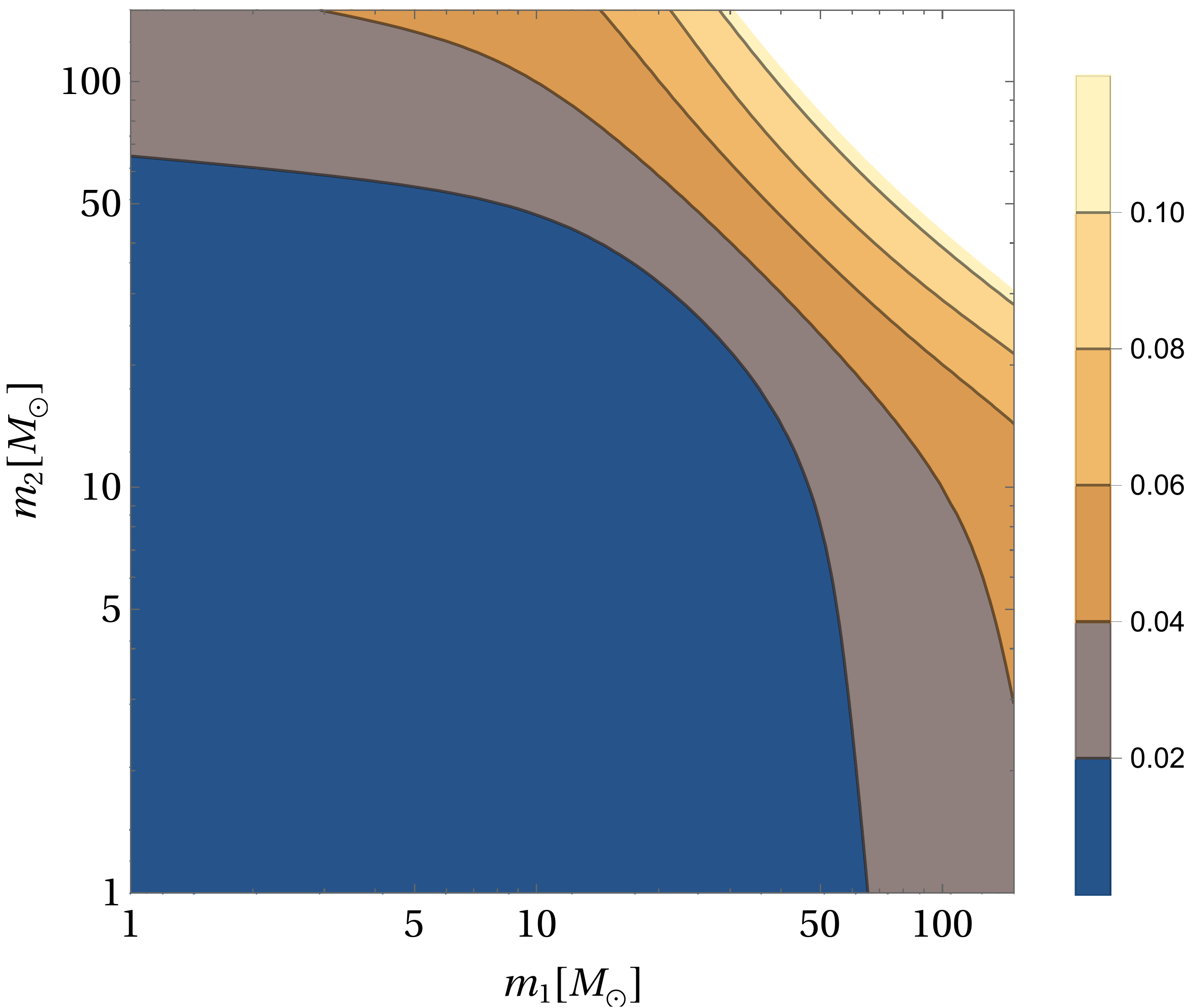}
	\caption{\label{ratio-log}The ratio of merger rate density from the second merger to that from the first merger, $\cR_2(t_0, m_1, m_2)/\cR_1(t_0, m_1, m_2)$, as a function of component masses for the log-normal mass function. We have fixed the hyper-parameters $\{M_c, \s, \fpbh\}$ to their best-fit values.}
\end{figure}

In \Fig{ratio-log}, we show the ratio of merger rate density from the second merger to the one from the first merger, namely $\cR_2(t_0, m_1, m_2)/\cR_1(t_0, m_1, m_2)$, by fixing the hyper-parameters $\{M_c, \s, \fpbh\}$ to their best-fit values. It can be seen that the second merger provides more contribution to the total merger rate density as component mass increases. Even though $\cR_2(t_0, m_1, m_2)/\cR_1(t_0, m_1, m_2)$ can reach as high as $\gtrsim 10\%$, the ratio of merger rate from second merger to that from the first merger is $R_2(t_0)/R_1(t_0) = 1.0^{+0.2}_{-0.1}\%$ and is negligible. This is because the major contribution to the merger rate is from the masses less than $50\Msun$, and the correction is negligible in this mass range. Therefore the effect of merger history can be safely ignored when estimating the merger rate of PBH binaries.

\subsection{Power-law mass function}

We next consider a PBH mass function with the power-law form of \cite{Carr:1975qj}
\e\label{power} 
P(m) = \frac{\al-1}{\Mmin} \(\frac{m}{\Mmin}\)^{-\al},
\q
where $\Mmin$ is the lower-mass cut-off such that $m>\Mmin$, and $\al>1$ is the power-law index. 
The power-law mass function can typically result from a broad or flat power spectrum of the curvature perturbations \cite{DeLuca:2020ioi} during radiation-dominated era \cite{Carr:2016drx,Carr:2017jsz}.
The hyper-parameters are $\Ld = \{\Mmin, \al, \fpbh\}$ in this case. 
We can then derive the averaged PBH mass and averaged number density from \Eq{mpbh} and \Eq{Fm} as
\e
\mpbh = \Mmin \frac{\al}{\al-1},
\q
\e 
F(m) = \frac{\al}{m} \(\frac{m}{\Mmin}\)^{-\al}.
\q

\begin{figure}[tbp!]
	\centering
	\includegraphics[width=0.5\textwidth]{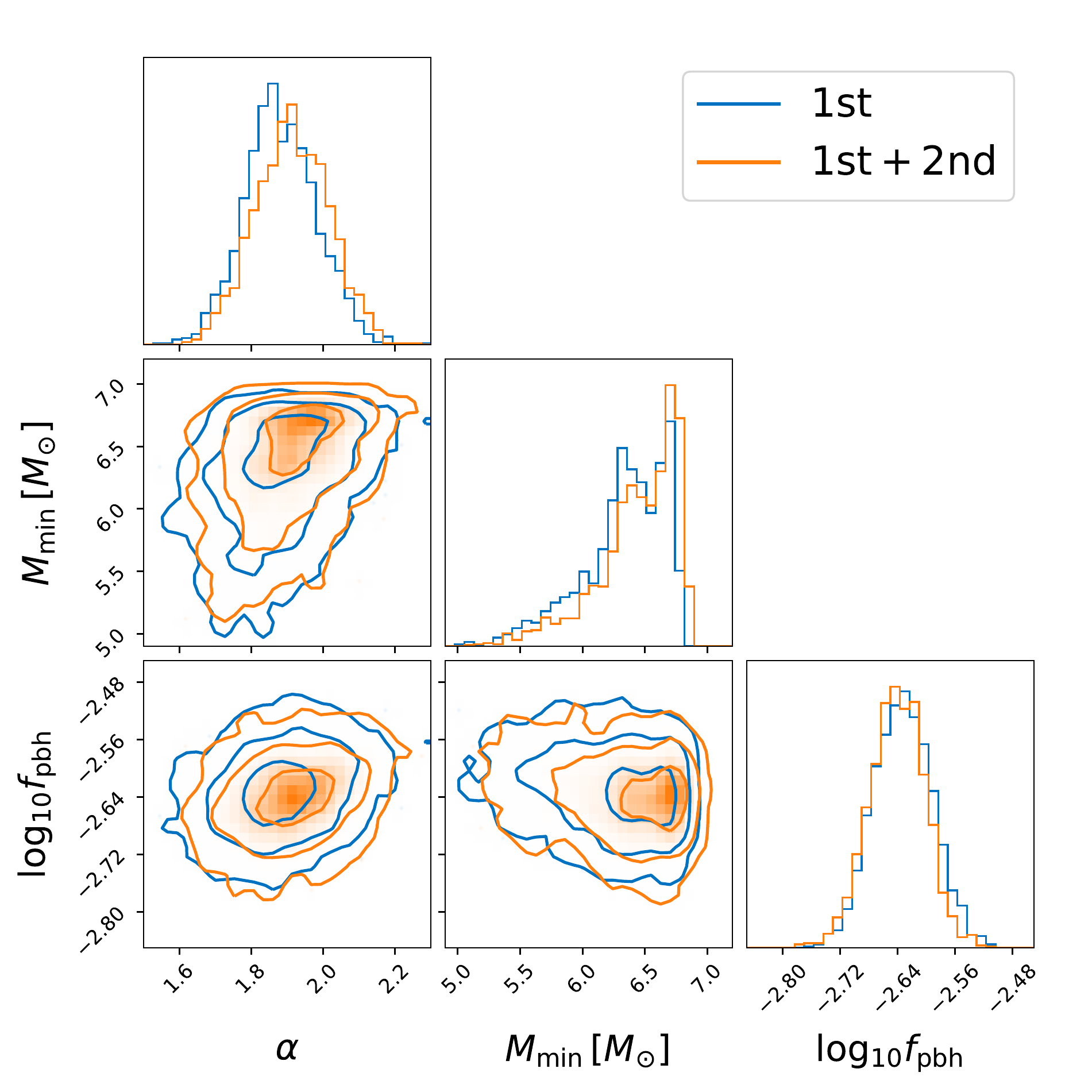}
	\caption{\label{posterior-power}The marginalized one- and two-dimensional posterior distributions for hyper-parameters $\{\Mmin, \al, \fpbh\}$ in the power-law mass function inferred from GWTC-3. The blue color denotes the results from the first merger only, while the orange denotes the results from both the first and second mergers. The contours represent the $1\sigma$, $2\sigma$, and $3\sigma$ credible regions, respectively.}
\end{figure}

Using $69$ BBHs from GWTC-3 and performing the hierarchical Bayesian inference, we obtain $\Mmin = 6.5^{+0.3}_{-0.8} \Msun$, $\al = 1.9^{+0.2}_{-0.2}$, and $\fpbh = 2.3^{+0.3}_{-0.3} \times 10^{-3}$. The posteriors for the hyper-parameters $\Ld = \{\Mmin, \al, \fpbh\}$ are shown in \Fig{posterior-power}. Note that we get a smaller value of $\al$ than that inferred from GWTC-1 in Ref.~\cite{Wu:2020drm} because GWTC-3 contains heavier BHs than those from GWTC-1. From \Eq{Rt}, we also infer the local merger rate as $R(t_0) = 48^{+15}_{-12} \gpcyr$. The results of the local merger rate and abundance of PBHs are consistent with the previous estimations \cite{Sasaki:2016jop,Ali-Haimoud:2017rtz,Chen:2018czv,Chen:2018rzo,Chen:2019irf,Wu:2020drm,Chen:2021nxo,Chen:2022fda,Zheng:2022wqo}, confirming that CDM cannot be dominated by the stellar-mass PBHs.

\begin{figure}[htbp!]
	\centering
	\includegraphics[width=0.5\textwidth]{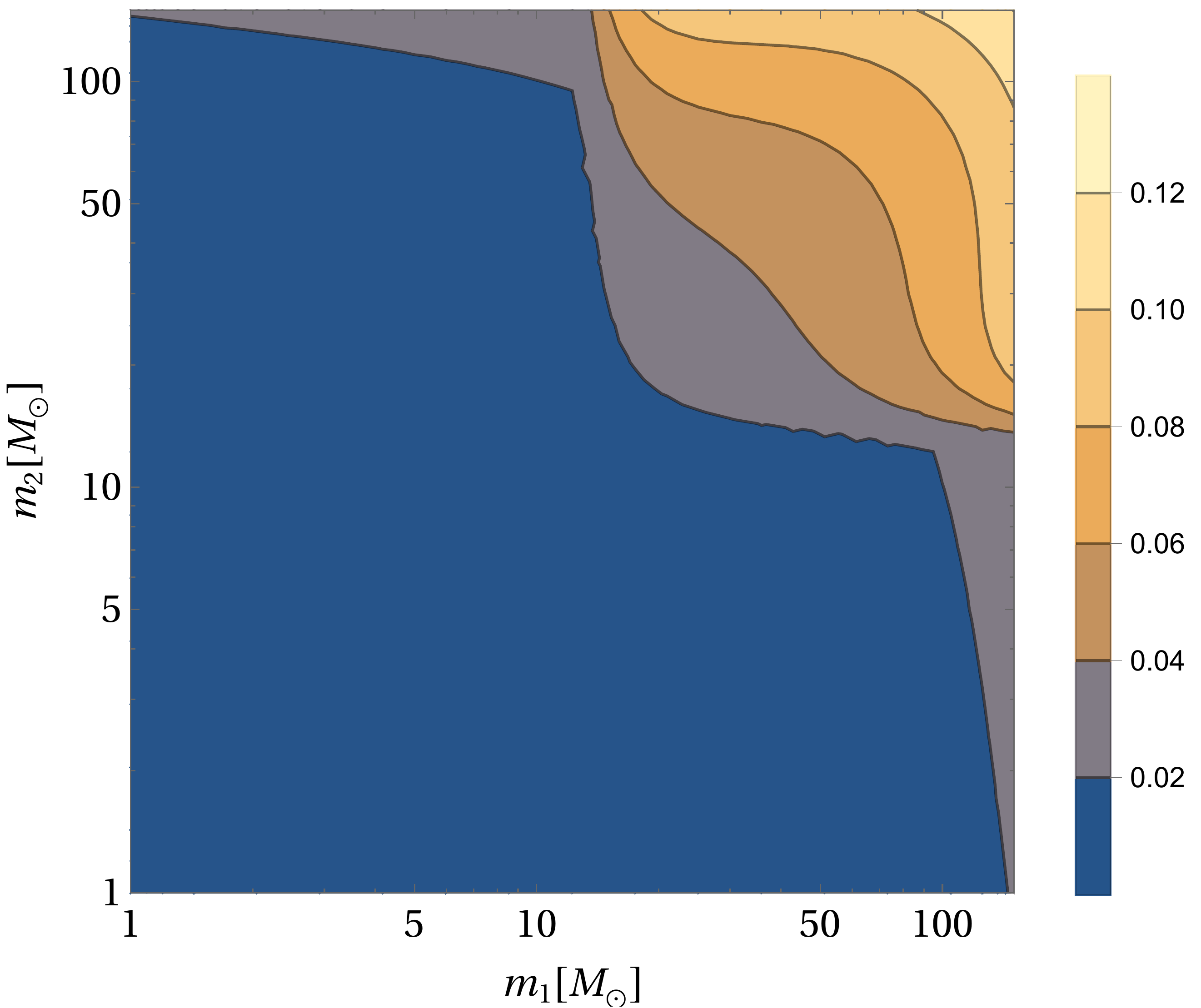}
	\caption{\label{ratio-power}The ratio of merger rate density from the second merger to that from the first merger, $\cR_2(t_0, m_1, m_2)/\cR_1(t_0, m_1, m_2)$, as a function of component masses for the power-law mass function. We have fixed the hyper-parameters $\{\Mmin, \al, \fpbh\}$ to their best-fit values.}
\end{figure}

In \Fig{ratio-power}, we show the ratio of merger rate density from the second merger to the one from the first merger, namely $\cR_2(t_0, m_1, m_2)/\cR_1(t_0, m_1, m_2)$, by fixing the hyper-parameters $\{\Mmin, \al, \fpbh\}$ to their best-fit values. It can be seen that the second merger provides more contribution to the total merger rate density as component mass increases. Even though $\cR_2(t_0, m_1, m_2)/\cR_1(t_0, m_1, m_2)$ can reach as high as $\gtrsim 10\%$, the ratio of merger rate from second merger to that from the first merger is $R_2(t_0)/R_1(t_0) = 0.9^{+0.1}_{-0.1}\%$ and is negligible. This is because the major contribution to the merger rate is from the masses less than $50\Msun$, and the correction is negligible in this mass range. Therefore the effect of merger history can be safely ignored when estimating the merger rate of PBH binaries.

\subsection{Broken power-law mass function}
We then consider a PBH mass function with the broken power-law form of \cite{Deng:2021ezy}
\begin{equation}
	P(m)= \left(\frac{m_*}{\alpha_1+1} + \frac{m_*}{\alpha_2-1}\right)^{-1} \begin{cases} (\frac{m}{m_*})^{\alpha_1}, & m<m_* \\ (\frac{m}{m_*})^{-\alpha_2}, & m>m_*\end{cases},
\end{equation}
where $m_*$ is the peak mass of $m P(m)$. Here $\al_1>0$ and $\al_2>1$ are two power-law indices. The broken power-law mass function is a generalization of the power-law form. It can be achieved if PBHs are formed by vacuum bubbles that nucleate during inflation via quantum tunneling \cite{Deng:2021ezy}.
The hyper-parameters are $\Ld = \{m_*, \al_1, \al_2, \fpbh\}$ in this case. 
We can then derive the averaged PBH mass and averaged number density from \Eq{mpbh} and \Eq{Fm} as
\e
\mpbh = \frac{\al_1 \al_2}{\(\al_1 +1\)\(\al_2 -1\)} m_*,
\q
\e 
F(m) = \frac{\al_1 \al_2}{\al_1 + \al_2} \begin{cases} (\frac{m}{m_*})^{\alpha_1}, & m<m_* \\ (\frac{m}{m_*})^{-\alpha_2}, & m>m_*\end{cases}.
\q

\begin{figure}[tbp!]
	\centering
	\includegraphics[width=0.5\textwidth]{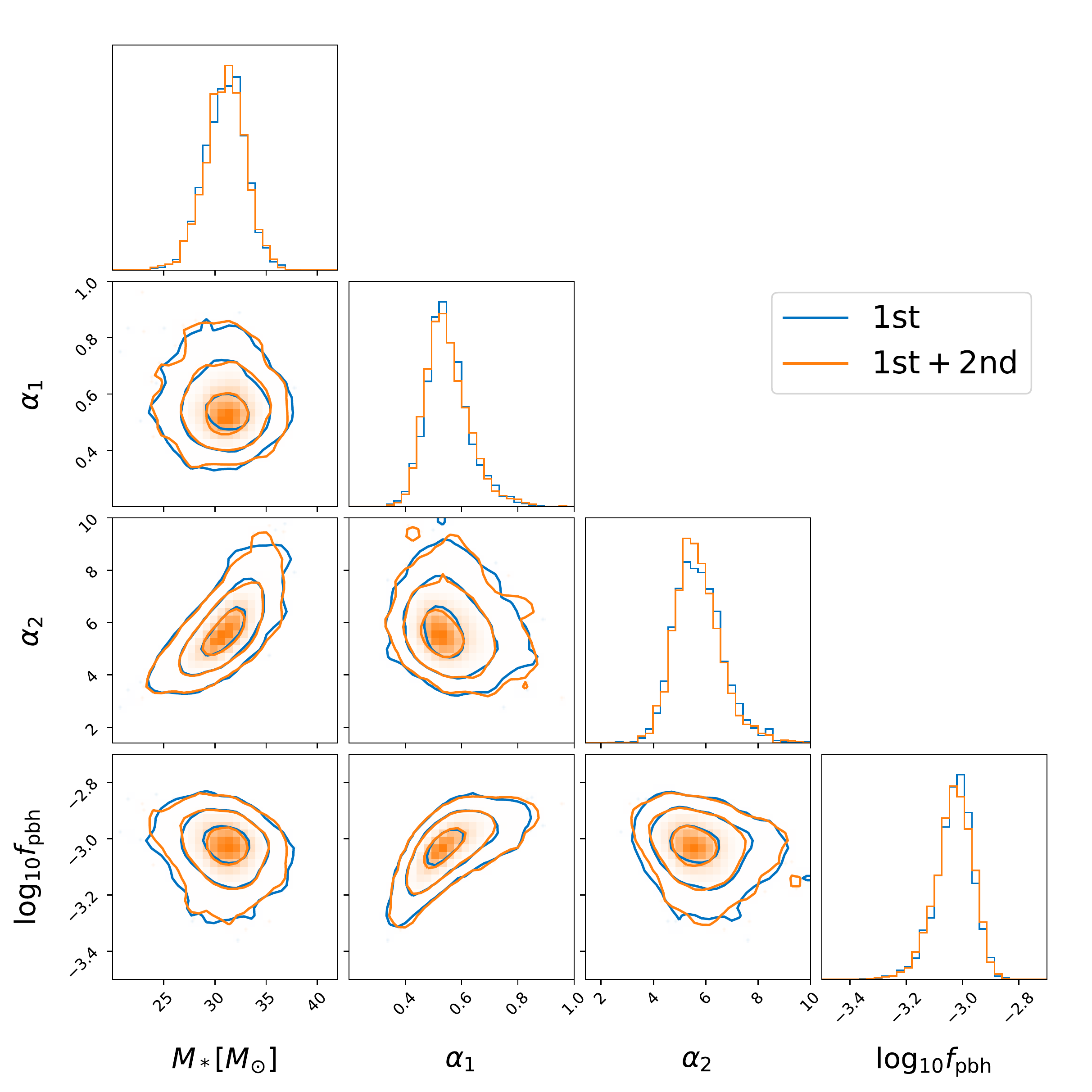}
	\caption{\label{posterior-bpower}The marginalized one- and two-dimensional posterior distributions for hyper-parameters $\{m_*, \al_1, \al_2, \fpbh\}$ in the broken power-law mass function inferred from GWTC-3. The blue color denotes the results from the first merger only, while the orange denotes the results from both the first and second mergers. The contours represent the $1\sigma$, $2\sigma$, and $3\sigma$ credible regions, respectively.}
\end{figure}

Using $69$ BBHs from GWTC-3 and performing the hierarchical Bayesian inference, we obtain $m_* = 31.1^{+1.8}_{-2.1} \Msun$, $\al_1 = 0.54^{+0.08}_{-0.06}$, $\al_2 = 5.6^{+0.9}_{-0.8}$, and $\fpbh = 0.9^{+0.1}_{-0.1} \times 10^{-3}$. The posteriors for the hyper-parameters $\Ld = \{m_*, \al_1, \al_2, \fpbh\}$ are shown in \Fig{posterior-bpower}. From \Eq{Rt}, we also infer the local merger rate as $R(t_0) = 46^{+15}_{-11} \gpcyr$. The results of the local merger rate and abundance of PBHs are consistent with the previous estimations \cite{Sasaki:2016jop,Ali-Haimoud:2017rtz,Chen:2018czv,Chen:2018rzo,Chen:2019irf,Wu:2020drm,Chen:2021nxo,Chen:2022fda,Zheng:2022wqo}, confirming that CDM cannot be dominated by the stellar-mass PBHs. We also confirm that there is a mass peak at $m_* \sim 34 \Msun$ as was found in  Ref.~\cite{Deng:2021ezy}.


\begin{figure}[tbp!]
	\centering
	\includegraphics[width=0.5\textwidth]{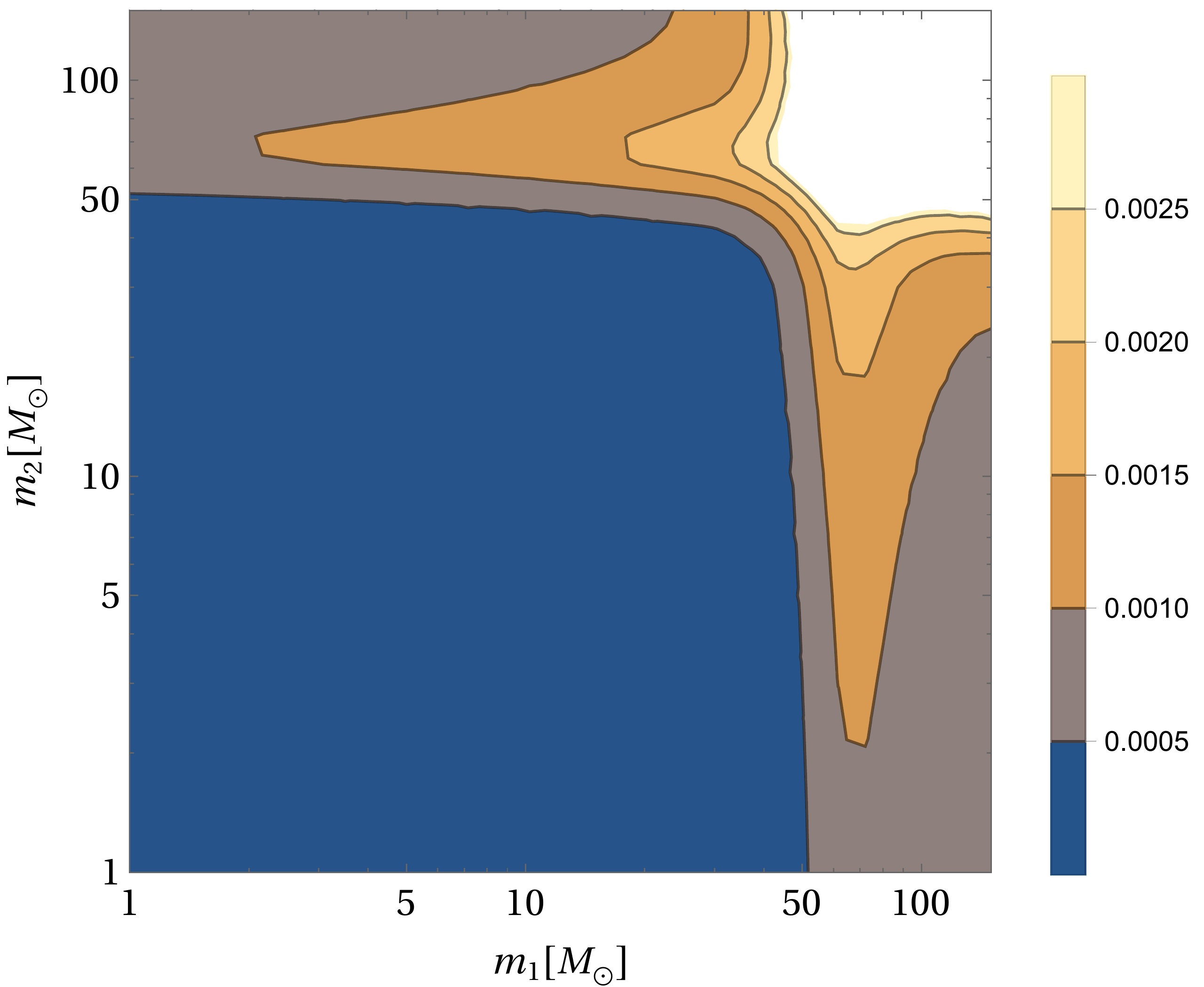}
	\caption{\label{ratio-bpower}The ratio of merger rate density from the second merger to that from the first merger, $\cR_2(t_0, m_1, m_2)/\cR_1(t_0, m_1, m_2)$, as a function of component masses for the broken power-law mass function. We have fixed the hyper-parameters $\{m_*, \al_1, \al_2, \fpbh\}$ to their best-fit values.}
\end{figure}

In \Fig{ratio-bpower}, we show the ratio of merger rate density from the second merger to the one from the first merger, namely $\cR_2(t_0, m_1, m_2)/\cR_1(t_0, m_1, m_2)$, by fixing the hyper-parameters $\{m_*, \al_1, \al_2, \fpbh\}$ to their best-fit values. It can be seen that the second merger provides more contribution to the total merger rate density as component mass increases. Even though $\cR_2(t_0, m_1, m_2)/\cR_1(t_0, m_1, m_2)$ can reach as high as $\gtrsim 10\%$, the ratio of merger rate from second merger to that from the first merger is $R_2(t_0)/R_1(t_0) = 0.9^{+0.3}_{-0.1}\%$ and is negligible. This is because the major contribution to the merger rate is from the masses less than $50\Msun$, and the correction is negligible in this mass range. Therefore the effect of merger history can be safely ignored when estimating the merger rate of PBH binaries.

\subsection{Critical collapse mass function}

We last consider a PBH mass function with the critical collapse form of \cite{Niemeyer:1997mt,Yokoyama:1998xd,Carr:2016hva,Gow:2020cou}
\e
P(m)=\frac{\alpha^2\,  m^\alpha}{\Mf^{1+\alpha}\, \Gamma(1 / \alpha)} \exp \left(-(m/\Mf)^{\alpha}\right),
\q
where $\alpha$ is a universal exponent relating to the critical collapse of radiation, and $M_{\mathrm{f}}$ is the mass scale at the order of horizon mass at the collapse epoch \cite{Carr:2016hva}. 
There is no lower mass cut-off for this mass spectrum, but it is exponentially suppressed above the mass scale of $\Mf$.
The critical collapse mass function is closely associated with a $\delta$-function power spectrum of the density fluctuations \cite{Niemeyer:1997mt,Yokoyama:1998xd,Carr:2016hva,Gow:2020cou}. 
The hyper-parameters are $\Ld = \{\Mf, \al, \fpbh\}$ in this case. 
We can then derive the averaged PBH mass and averaged number density from \Eq{mpbh} and \Eq{Fm} as
\e
\mpbh = \frac{\Mf \Gamma(1 / \alpha)}{\alpha},
\q
\e 
F(m) = \al \Mf^{-\al} m^{\al -1} \exp\(-(m/\Mf)^{\al}\).
\q

\begin{figure}[tbp!]
	\centering
	\includegraphics[width=0.5\textwidth]{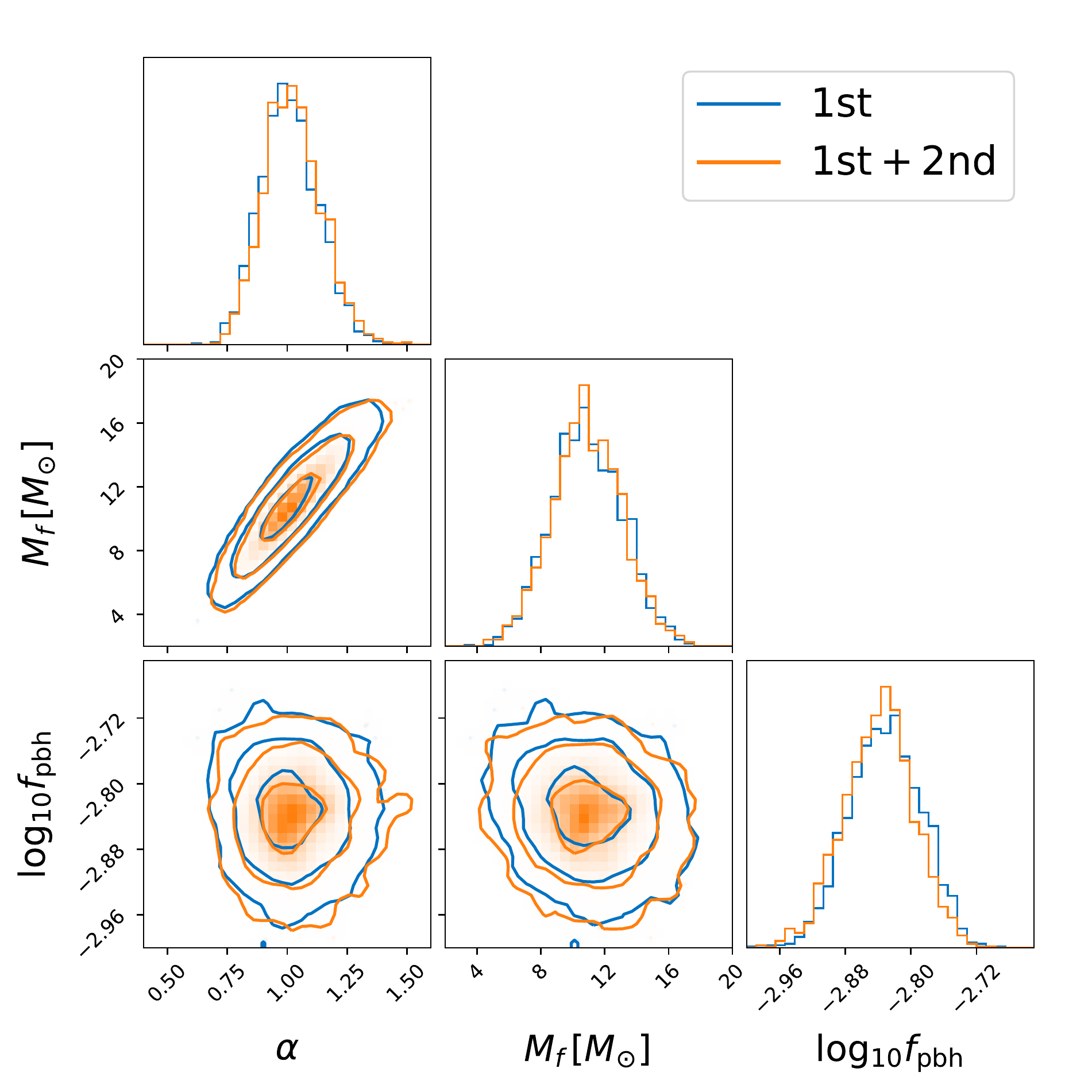}
	\caption{\label{posterior-CC}The marginalized one- and two-dimensional posterior distributions for hyper-parameters $\{\Mf, \al, \fpbh\}$ in the critical collapse mass function inferred from GWTC-3. The blue color denotes the results from the first merger only, while the orange denotes the results from both the first and second mergers. The contours represent the $1\sigma$, $2\sigma$ and $3\sigma$ credible regions, respectively.}
\end{figure}

Using $69$ BBHs from GWTC-3 and performing the hierarchical Bayesian inference, we obtain $\Mf = 10.8^{+3.7}_{-3.6} \Msun$, $\al = 1.0^{+0.2}_{-0.2}$, and $\fpbh = 1.5^{+0.2}_{-0.2} \times 10^{-3}$. The posteriors for the hyper-parameters $\Ld = \{\Mf, \al, \fpbh\}$ are shown in \Fig{posterior-CC}. From \Eq{Rt}, we also infer the local merger rate as $R(t_0) = 49^{+26}_{-16} \gpcyr$. The results of the local merger rate and abundance of PBHs are consistent with the previous estimations \cite{Sasaki:2016jop,Ali-Haimoud:2017rtz,Chen:2018czv,Chen:2018rzo,Chen:2019irf,Wu:2020drm,Chen:2021nxo,Chen:2022fda,Zheng:2022wqo}, confirming that CDM cannot be dominated by the stellar-mass PBHs.

\begin{figure}[tbp!]
	\centering
	\includegraphics[width=0.5\textwidth]{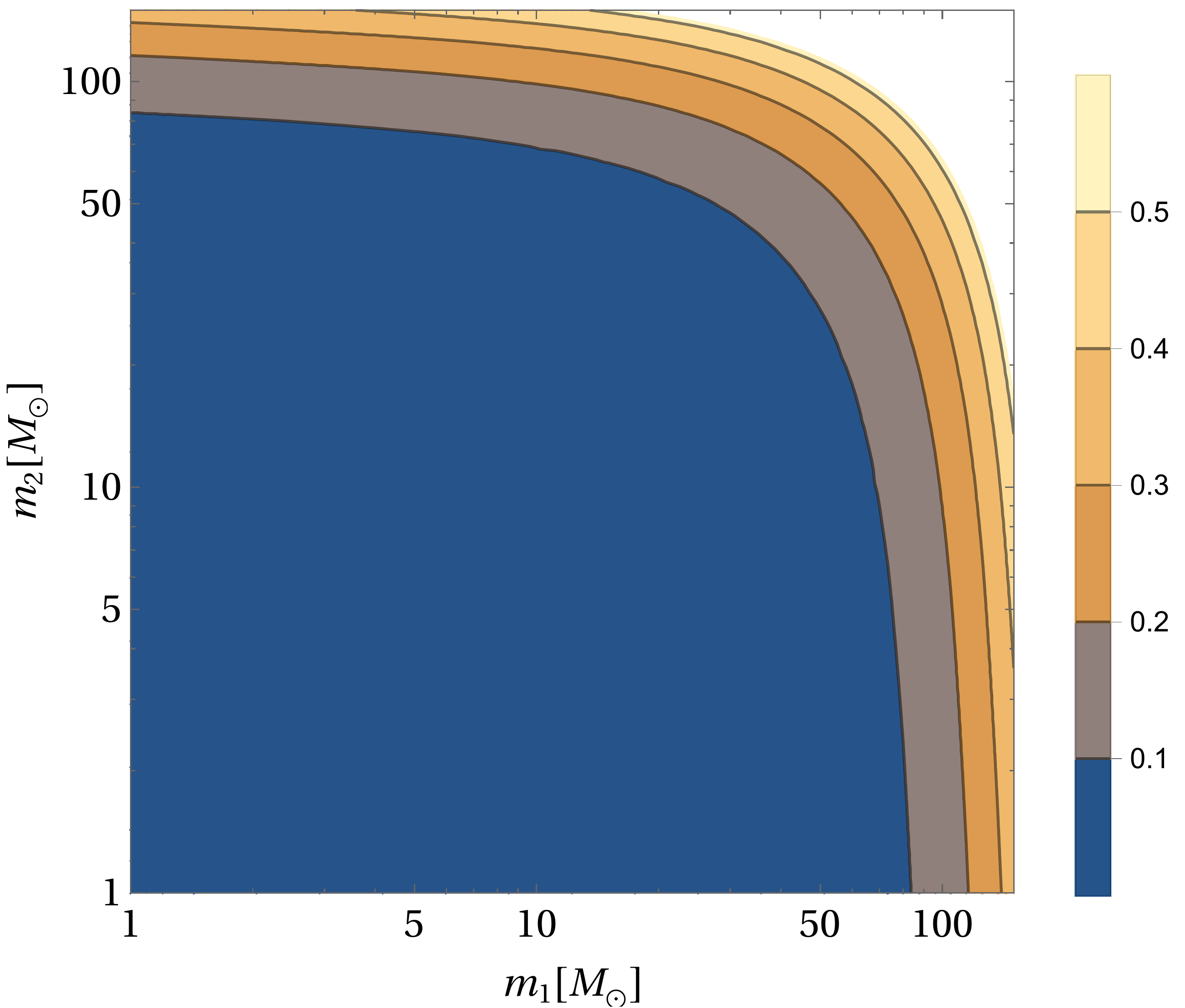}
	\caption{\label{ratio-CC}The ratio of merger rate density from the second merger to that from the first merger, $\cR_2(t_0, m_1, m_2)/\cR_1(t_0, m_1, m_2)$, as a function of component masses for the critical collapse mass function. We have fixed the hyper-parameters $\{\Mf, \al, \fpbh\}$ to their best-fit values.}
\end{figure}

In \Fig{ratio-CC}, we show the ratio of merger rate density from the second merger to the one from the first merger, namely $\cR_2(t_0, m_1, m_2)/\cR_1(t_0, m_1, m_2)$, by fixing the hyper-parameters $\{\Mf, \al, \fpbh\}$ to their best-fit values. It can be seen that the second merger provides more contribution to the total merger rate density as component mass increases. Even though $\cR_2(t_0, m_1, m_2)/\cR_1(t_0, m_1, m_2)$ can reach as high as $\gtrsim 10\%$, the ratio of merger rate from second merger to that from the first merger is $R_2(t_0)/R_1(t_0) = 2.2^{+1.3}_{-0.1}\%$ and is negligible. This is because the major contribution to the merger rate is from the masses less than $50\Msun$, and the correction is negligible in this mass range. Therefore the effect of merger history can be safely ignored when estimating the merger rate of PBH binaries.

\section{\label{conclusion}Conclusion}

In this work, we use $69$ BBHs from GWTC-3 to constrain the merger history of PBH binaries by assuming the observed BBHs from LVK are attributed to PBHs. We perform comprehensive Bayesian analyses by considering four commonly used PBH mass functions in literature, namely the log-normal, power-law, broken power-law, and critical collapse mass functions. 

We summarize the key results in \Table{table:BF}. 
It can be seen that the contribution of the merger rate from the second merger to the total merger rate is less than $5\%$.
Therefore, the higher-order hierarchical merger after the first one has a subdominant effect, and this effect can be neglected when evaluating the merger rate of PBH binaries.
It can also be seen that the Bayes factors for the model with a second merger versus the model with only the first merger, $\mathrm{BF}^{\mathrm{2nd}}_{\mathrm{1st}}$, are all smaller than $3$, indicating the evidence for the second merger is ``not worth more than a bare mention" \cite{BF}. In this sense, the Bayes factors also imply that the effect of merger history can be ignored.

Furthermore, for all four mass functions, we infer the abundance of PBH in CDM, $\fpbh$, to be at the order of $\mathcal{O}(10^{-3})$. The results of the local merger rate and abundance of PBHs are consistent with the previous estimations \cite{Sasaki:2016jop,Ali-Haimoud:2017rtz,Chen:2018czv,Chen:2018rzo,Chen:2019irf,Wu:2020drm,Chen:2021nxo,Chen:2022fda,Zheng:2022wqo}, confirming that CDM cannot be dominated by the stellar-mass PBHs. PBHs cluster at the late time of the Universe may play an important role in the merger rate. For all of the four PBH mass functions, we always have $f_{\mathrm{pbh}} \lesssim 3\times10^{-3}$. Therefore, according to Ref.~\cite{Hutsi:2020sol}, this effect can be safely ignored.

\begin{table}[tbp!]
	\centering
	\begin{tabular}{l|c|c|c|c}
		\hline\hline
		& LN & PL & BPL & CC \\
		\hline
		$\mathrm{BF}^{\mathrm{2nd}}_{\mathrm{1st}}$ & $0.9$ & $0.4$ & $0.89$ & $1.2$ \\
		$\mathrm{BF}_{\mathrm{PL}}$ & $166$ & $1$ & $49$ & $139$ \\
		$10^{3}\fpbh$ & $1.8^{+0.3}_{-0.3}$ & $2.3^{+0.3}_{-0.3}$ & $0.9^{+0.1}_{-0.1}$ & $1.5^{+0.2}_{-0.2}$ \\
		$10^{2}R_2/R_1$ & $1.0^{+0.2}_{-0.1}$ & $0.9^{+0.1}_{-0.1}$ & $1.3^{+0.3}_{-0.1}$ & $2.2^{+1.3}_{-0.5}$ \\
		\hline
	\end{tabular}
	\caption{\label{table:BF}Summary of the key results for the log-normal (LN), power-law (PL), broken power-law (BPL), and critical collapse (CC) mass functions. The first row, $\mathrm{BF}^{\mathrm{2nd}}_{\mathrm{1st}}$, shows the Bayes factors for the model with 2nd merger versus the model with only 1st merger; the second row, $\mathrm{BF}_{\mathrm{PL}}$, shows the Bayes factors for the model with different PBH mass function versus the model with the power-law PBH mass function by accounting for the second merger effect; the third row, $\fpbh$, shows the abundance of PBH in CDM inferred from different models by accounting for the second merger effect; the last row, $R_2/R_1$, shows the merger rate ratio between the second merger and the first merger.}
\end{table}

We also compute the Bayes factors between the models with different PBH mass functions. The Bayes factors $\mathrm{BF}_{\mathrm{PL}}$ are estimated by taking the model with the power-law mass function as the fiducial model. 
We find that $\mathrm{BF}_{\mathrm{PL}}^{\mathrm{LG}}$ has the largest value, indicating that the log-normal mass function can best fit GWTC-3 among the four mass functions considered in this work. Our findings contradict the results from Ref.~\cite{Deng:2021ezy} claiming that the broken power-law mass function can fit better than the log-normal form. There are some drawbacks from analyses in Ref.~\cite{Deng:2021ezy}. Firstly, Ref.~\cite{Deng:2021ezy} neglects the uncertainties in measuring each event's masses and completely ignores the redshift evolution of the merger rate. Secondly, Ref.~\cite{Deng:2021ezy} deals with the selection effect of GW detectors improperly. In this sense, we disagree with Ref.~\cite{Deng:2021ezy} and conclude that the most frequently used log-normal mass function can fit GWTC-3 best among the four mass functions.

\begin{acknowledgments}
We thank the referee for providing constructive comments and suggestions to improve the quality of this paper. 
We would also like to thank Xingjiang Zhu, Xiao-Jin Liu, Shen-Shi Du, and Zhu Yi for the useful discussions.
ZCC is supported by the National Natural Science Foundation of China (Grant No.~12247176 and No.~12247112) and the China Postdoctoral Science Foundation Fellowship No.~2022M710429.
ZQY is supported by the China Postdoctoral Science Foundation Fellowship No.~2022M720482.
LL is supported by the National Natural Science Foundation of China (Grant No.~12247112 and No.~12247176).

This research has made use of data, software and/or web tools obtained from the Gravitational Wave Open Science Center (\url{https://www.gw-openscience.org/}), a service of LIGO Laboratory, the LIGO Scientific Collaboration, the Virgo Collaboration, and the KAGRA Collaboration. 
LIGO Laboratory and Advanced LIGO are funded by the United States National Science Foundation (NSF) as well as the Science and Technology Facilities Council (STFC) of the United Kingdom, the Max-Planck-Society (MPS), and the State of Niedersachsen/Germany for support of the construction of Advanced LIGO and construction and operation of the GEO600 detector. Additional support for Advanced LIGO was provided by the Australian Research Council. Virgo is funded, through the European Gravitational Observatory (EGO), by the French Centre National de Recherche Scientifique (CNRS), the Italian Istituto Nazionale di Fisica Nucleare (INFN) and the Dutch Nikhef, with contributions by institutions from Belgium, Germany, Greece, Hungary, Ireland, Japan, Monaco, Poland, Portugal, Spain. The construction and operation of KAGRA are funded by Ministry of Education, Culture, Sports, Science and Technology (MEXT), and Japan Society for the Promotion of Science (JSPS), National Research Foundation (NRF) and Ministry of Science and ICT (MSIT) in Korea, Academia Sinica (AS) and the Ministry of Science and Technology (MoST) in Taiwan.
\end{acknowledgments}	
	
\bibliography{ref}

\end{document}